\documentclass[12pt]{article}
\usepackage{epsfig}
\usepackage{graphicx}
\usepackage{subfigure}
\usepackage{amssymb}
\usepackage{pifont}
\usepackage{latexsym}

\usepackage{amsmath}
\usepackage{amsfonts}
\usepackage{fancyhdr}
\usepackage{amscd}

\title{Mode propagation and attenuation in lined ducts}
\author{WenPing Bi, Vincent Pagneux}

\begin{document}
\maketitle
\begin{center}
Laboratoire d'Acoustique de l'Universit\'e du Maine,\\
UMR CNRS 6613\\
Av. O Messiaen, 72085 LE MANS Cedex 9,\\
France\\
\end{center}

\begin{abstract}
Optimal impedance for each mode is an important concept in an infinitely long duct lined with uniform absorption material. However it is not valid for finite length linings. This is because that the modes in lined ducts are not power-orthogonal; the total sound power is not equal to the sum of the sound power of each mode; cross-power terms may play important roles. In this paper, we study sound propagation and attenuation in an infinite rigid duct lined with a finite length of lining impedance. The lining impedance may be axial segments and circumferentially non-uniform. We propose two new  physical quantities $\mathsf{K_p}$ and $\mathsf{S}$ to describe the self-overlap of the left eigenfunction and right eigenfunction of one mode and the normalized overlap between modes, respectively. The two new physical quantities describe totally the mode behaviors in lined ducts.
\end{abstract}

\maketitle

\section{Introduction}

Optimal impedance is an important concept in lined ducts. It provides a theoretical limit for the maximum attenuation that can be obtained. The concept of optimal impedance was proposed firstly by Cremer\cite{cremer} for the so called ``least attenuation mode" (usually the plane mode). It was extended by Tester\cite{tester1, tester2} for any guided modes and shown to correspond to the double roots of the dispersion equation. Similar conclusion was also proposed by Shenderov\cite{shenderov}. However, this concept is only correct in an infinitely lined duct, which is never the case in practice.

In a lined duct,  length of lining is finite, the lining impedance maybe uniform or non-uniform along axial or transverse directions. In this case, the concept of optimal impedance is not valid. This optimum impedance might not exist. A number of numerical studies were proposed to try to increase the attenuation by parameter studies\cite{unruh, watson, fuller, regan, mani, bi, tam}. However, to the best of our knowledge, we do not know, till now, what physical quantities play important roles in the attenuation of sound power in a practical lined duct.

Mode scattering is expected as the main mechanism to increase the attenuation. But sometimes, this scattering effects are impressive, e.g., the penalty effects of splices\cite{bi, tam, tester, boden}; others, they might be not visible\cite{rademaker}. We still have not a deep understanding about mode scattering, specially the liner mode scattering.

Modes in lined ducts are not orthogonal in the standard definition 
\begin{equation}\label{standard}
\int_s\tilde{\phi}_i^*\tilde{\phi}_jds=\Lambda_i\delta_{ij},
\end{equation}
as in rigid or non-absorptive ducts where left eigenfunctions $\tilde{\phi}_i^l$ are equal to right eigenfunctions $\tilde{\phi}_i^r$, $\tilde{\phi}_i^l=\tilde{\phi}_i^r=\tilde{\phi}_i$, ''*'' refers to complex conjugate, $\Lambda_i$ refers to normalization constant. When the lining impedance is absorptive, or in other words, it is complex, left eigenfunctions are not equal to right eigenfunctions. Modes are bi-orthogonal
\begin{equation}\label{modify}
\int_s(\tilde{\phi}_i^l)^*\tilde{\phi}_j^rds=\Lambda_i\delta_{ij}.
\end{equation}
If there exist some kinds of symmetric, left eigenfunctions may be equal to the conjugate of right eigenfunctions i.e., $(\tilde{\phi}_i^l)^*=\tilde{\phi}_i^r$. In this case, the bi-orthogonal condition is $\int_s\tilde{\phi}_i^r\tilde{\phi}_j^rds=\Lambda_i\delta_{ij}$. When modes are not orthogonal, the total sound power is not equal to the sum of the sound power of each mode; the cross-power terms may play important roles. Therefore, although the attenuation for each mode might be maximum near the corresponding optimum impedance, the total sound power is not optimum because of the cross-power.   

In this paper, we study sound propagation in a uniformly rigid duct fitted with finite length of lining without flow. We express the sound field in the lined region in terms of liner modes. We define two new physical quantities $\mathsf{K_p}$ and $\mathsf{S}$ to describe mode self-overlap and mutual overlap, respectively, in the lined region. We show that these two new physical quantities play essentially roles in the sound attenuation in lined region.

\section{Derivation of Equations}

\begin{figure}
\begin{center}
\includegraphics[width=90mm]{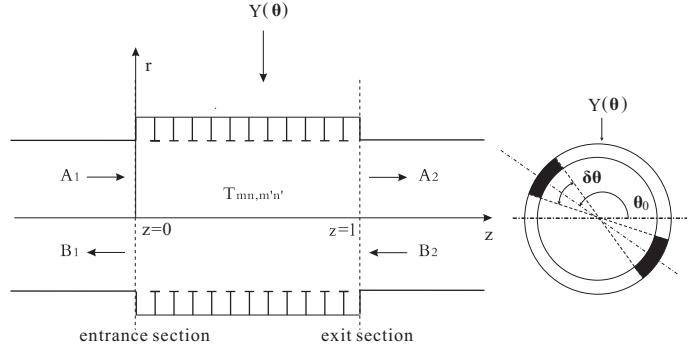}
\end{center}
\caption{\label{config1} Configuration of one axial segment
  nonuniform lined duct.}
\end{figure}

We consider an infinite rigid duct with circular cross section lined with a
region of circumferentially nonuniform liner. The liner properties are assumed to
be given by a distribution of locally reacting impedance. Without
significant loss of generality, the distribution may be assumed
axially segmented, i.e., the impedance is set piecewise constant
along the duct, while being arbitrarily variable along the
circumference of each segment. In Fig.~\ref{config1} the
configuration of one axial segment of lining impedance is
depicted, the circumferential variation of impedance is presented
as two acoustically rigid splices, which is a typical
configuration in the intake of an aeroengine. For simplicity, we assume the circumferential nonuniformity has a mirror symmetry. Linear and lossless sound
propagation in air is assumed. With time
dependence $\exp(j\omega t)$ omitted, the equation of mass
conservation combined with the equation of state, and the equation
 of momentum conservation are written as
\begin{eqnarray}
\mathbf{\nabla}\cdot\mathbf{v}&=&-\frac{j\omega}{\rho_0 c_0^2} p,\label{mass}\\
j\omega\mathbf{v}&=&-\frac{1}{\rho_0}\mathbf{\nabla}
p,\label{momentum}
\end{eqnarray}
where $\mathbf{v}$ is the particle velocity, $p$ is the acoustic
pressure, and  $\rho_0$ and $c_0$ are the ambient density and
speed of sound in air. Pressures, velocities and lengths are
respectively divided by $\rho_0 c_0^2$, $c_0$ and $R$ (the duct
radius) to reduce Eqs. (\ref{mass})-(\ref{momentum}) to the
dimensionless form
\begin{eqnarray}
& &\mathbf{\nabla}\cdot\mathbf{v} =-jKp,\label{euler1}\\
& &-jK\mathbf{v}=\mathbf{\nabla} p,\label{euler2}
\end{eqnarray}
where $K=\omega R /c_0$ is the dimensionless wave number. This
yields the three dimensional wave equation

\begin{equation}\label{helmholtz}
\nabla^2_\bot p+\frac{\partial^2p}{\partial z^2}+K^2p=0,
\end{equation}
where
\begin{equation}
\nabla^2_\bot =\frac{1}{r}\frac{\partial}{\partial
  r}(r\frac{\partial}{\partial
  r})+\frac{1}{r^2}\frac{\partial^2}{\partial \theta^2}.
\end{equation}

In the lined part, the radial boundary condition is
\begin{equation}\label{bound1}
\frac{\partial p}{\partial r}=Y(\theta)p, \; \;  \mathrm{at} \; \; r=1,
\end{equation}
where $Y(\theta)=-jK \beta(\theta)$, and $\beta(\theta)$ is the liner admittance. 

Sound pressure in lined part is expanded over right normalized eigenfunctions $\mathbf{\Phi}^r(r, \theta)$ of liner modes
\begin{equation}\label{solution}
p(r, \theta, z)=(\mathbf{\Phi}^r)^T\mathsf{E_2}(z)\mathbf{C'_1}+(\mathbf{\Phi}^r)^T\mathsf{E_2}^{-1}(z)\mathbf{C'_2},
\end{equation}
where $\mathbf{C'_1}$ and $\mathbf{C'_2}$ are amplitude vectors of dimension $N_{t}$, $\mathsf{E_2}(z)$ and $\mathsf{E_2^{-1}}(z)$ are
diagonal matrices with $\exp(-jK^z_iz)$ and $\exp(jK^z_iz)$ respectively on the main diagonal, with $K^z_i$ being the axial wavenumber of liner mode $i$, $"^T"$ refers to transpose. The right eigenfunction $\mathbf{\Phi}^r(r, \theta)$ of lined modes is calculated by expanding in term of rigid duct modes and an additional function that carries the information about the impedance boundary\cite{bi1, bi2}. $N_{t}=M\times N$, where $M$ and $N$ refer to the truncation in circumference and radial direction, respectively, of the expansion. The right eigenfunctions of lined modes are normalized in standard definition
\begin{equation}
\int(\tilde{\phi}^r_i)^*\tilde{\phi}^r_ids=\Lambda^r_i.
\end{equation}
It is noted that the right eigenfunctions are not orthogonal under the standard definition (\ref{standard}).

Before proceeding, it is convenient to rewrite Eq. (\ref{solution}) in the following form, by redefining the amplitude coefficients $\mathbf{C_1}$ and $\mathbf{C_2}$ 
\begin{equation}
p(r, \theta, z)=(\mathbf{\Phi}^r)^T\mathsf{E_2}(z-L_0)\mathbf{C_1}+(\mathbf{\Phi}^r)^T\mathsf{E_2}(L-z)\mathbf{C_2},\label{solutionnew}
\end{equation}
where $L_0$ and $L$ refer to the beginning and end of the lined part. Without loss of generality,  we assumed  in the followings $L_0=0$. In the form of Eq. (\ref{solutionnew}), numerical stability is ensured because the propagation matrices $\mathsf{E_2}(z-L_0)$ and $\mathsf{E_2}(L-z)$ have only positive arguments and contain no exponentially diverging terms due to the evanescent modes.

The continuity of  pressure and axial velocity leads to
\begin{eqnarray}
\mathbf{\Psi}^T(\mathbf{A} + \mathbf{B}) &=& (\mathbf{\Phi}^r)^T( \mathbf{C_1} +  \mathsf{E_{2}}(L)\mathbf{C_2}),\nonumber \\
\mathbf{\Psi}^T\mathsf{K_0}\left( \mathbf{A} - \mathbf{B}\right) &=& (\mathbf{\Phi}^r)^T\mathsf{K_Y} \left(\mathbf{C_1} - \mathsf{E_{2}}(L) \mathbf{C_2}\right),\label{eq:conti}\\
\mathbf{\Psi}^T\mathsf{E_{1}}\mathbf{D} &=& (\mathbf{\Phi}^r)^T\left( \mathsf{E_2}(L)\: \mathbf{C_1} + \mathbf{C_2}\right),\nonumber  \\
\mathbf{\Psi}^T\mathsf{K_0}\mathsf{E_{1}}\mathbf{D} &=& (\mathbf{\Phi}^r)^T\mathsf{K_Y}\left( \mathsf{E_2}(L) \: \mathbf{C_1} -
\mathbf{C_2}\right),\nonumber
\end{eqnarray}
where $\mathbf{\Psi}$ refers to the eigenfunctions of rigid modes, $\mathsf{E_{1}}$, $\mathsf{E_2}$ are diagonal matrices with $e^{-jK_{mn,rigid}^zL}$ and $e^{-jK_i^zL}$ on diagonal, respectively, $\mathsf{K_0}$ and $\mathsf{K_Y}$ are diagonal matrices with the axial wavenumbers on diagonal in the rigid and lined sections (resp. the $K_{mn, rigid}^z/K$ and $K_i^z/K$). 

Projecting the eq. (\ref{eq:conti}) over left normalized eigenfunctions $\mathbf{\Phi}^l$,
\begin{eqnarray}
\mathsf{F}(\mathbf{A} + \mathbf{B}) &=& \mathsf{E}( \mathbf{C_1} +  \mathsf{E_2}\mathbf{C_2}),\nonumber \\
\mathsf{F}\mathsf{K_0}\left( \mathbf{A} - \mathbf{B}\right) &=& \mathsf{E}\mathsf{K_Y} \left(\mathbf{C_1} - \mathsf{E_{2}} \mathbf{C_2}\right),\label{eq:conti1}\\
\mathsf{F}\mathsf{E_1}\mathbf{D} &=& \mathsf{E}\left( \mathsf{E_2}\: \mathbf{C_1} + \mathbf{C_2}\right),\nonumber  \\
\mathsf{F}\mathsf{K_0}\mathsf{E_{1}}\mathbf{D} &=& \mathsf{E}\mathsf{K_Y}\left( \mathsf{E_2} \: \mathbf{C_1} -
\mathbf{C_2}\right),\nonumber
\end{eqnarray}
where 
\begin{eqnarray}
\mathsf{F}=\int(\mathbf{\Phi}^l)^*\mathbf{\Psi}^Tds & \mathsf{E}=\int(\mathbf{\Phi}^l)^*(\mathbf{\Phi}^r)^Tds,
\end{eqnarray}
For simplicity, we have written $\mathsf{E_2}(L)$ and $\mathsf{E_1}(L)$ as $\mathsf{E_2}$ and $\mathsf{E_1}$, respectively. $\mathsf{F}$ describes the mode couplings among rigid modes and liner modes. Because we have assumed the circumferential nonuniformity has a mirror symmetry, we have $(\tilde{\phi}^l_i)^*=\tilde{\phi}^r_i$, $\Lambda_i^l=\Lambda_i^r$. The matrix $\mathsf{E}$ is diagonal, its elements in the main diagonal are  
\begin{equation}
E_{ii}=\frac{\int (\tilde{\phi}_i^l)^*\tilde{\phi}_i^rds}{\sqrt{\Lambda_i^l\Lambda_i^r}}=\frac{\int \tilde{\phi}_i^r\tilde{\phi}_i^rds}{\sqrt{\Lambda_i^r\Lambda_i^r}},
\end{equation}
where $\tilde{\phi}_i^l$ and $\tilde{\phi}_i^r$ are left and right eigenfunctions without normalization, $\Lambda_i^{l(r)}$ are the normalization constants for left and right eigenfunctions, respectively. From eq. (\ref{eq:conti1}), we obtain the liner mode amplitude 
\begin{eqnarray}
\mathbf{C_0} &=& (\mathsf{I}-\mathsf{G_2}\mathsf{G_1^{-1}}\mathsf{E_2}\mathsf{G_2}\mathsf{G_1}^{-1}\mathsf{E_2})^{-1}(\mathsf{G_1}-\mathsf{G_2}\mathsf{G_1}^{-1}\mathsf{G_2})\mathbf{A},\\
\mathbf{C_1} &=& \frac{1}{2}\mathsf{E}^{-1}\mathbf{C_0},\\
\mathbf{C_2} &=& \frac{1}{2}\mathsf{E}^{-1}\mathsf{G_3}\mathbf{C_0},
\end{eqnarray}
where
\begin{eqnarray}
\mathsf{G_1}=\mathsf{F}+\mathsf{K}_Y^{-1}\mathsf{F}\mathsf{K}_0, & \mathsf{G_2}=\mathsf{F}-\mathsf{K}_Y^{-1}\mathsf{F}\mathsf{K}_0, & \mathsf{G_3}=\mathsf{G_2}\mathsf{G_1}^{-1}\mathsf{E_2}.
\end{eqnarray}

Sound power in lined part,
\begin{eqnarray}\label{sp}
W &=& \frac{1}{2}\Re\{\int p(r, \theta)v_z^*(r, \theta)ds\}\\\nonumber
     &=& \frac{1}{2}\Re\{[(\mathbf{\Phi}^r)^T\mathsf{E_2}(z)\mathbf{C_1}+(\mathbf{\Phi}^r)^T\mathsf{E_2}(L-z)\mathbf{C_2}] \\\nonumber
     &  &\;\;\;\;\;\; \times[(\mathbf{\Phi}^r)^T\mathsf{K_Y}\mathsf{E_2}(z)\mathbf{C_1}-(\mathbf{\Phi}^r)^T\mathsf{K_Y}\mathsf{E_2}(L-z)\mathbf{C_2}]^*\}\\\nonumber
     &=& \frac{1}{8}\Re\{\mathbf{C_0}^T(\mathsf{K_p}\cdot\mathsf{S})\mathsf{K_Y}^*\mathsf{E_3}\mathbf{C_0}^* +  \mathbf{C_0}^T(\mathsf{K_p}\cdot\mathsf{S})\mathsf{K_Y}^*\mathsf{E_5}\mathsf{E_4}\mathbf{G_3}^*\mathbf{C_0}^*\\\nonumber
     & + & \mathbf{C_0}^T\mathbf{G_3}^T(\mathsf{K_p}\cdot\mathsf{S})\mathsf{K_Y}^*\mathsf{E_7}\mathsf{E_6}\mathbf{C_0}^* + \mathbf{C_0}^T\mathbf{G_3}^T(\mathsf{K_p}\cdot\mathsf{S})\mathsf{K_Y}^*\mathsf{E_8}\mathbf{G_3}^*\mathbf{C_0}^*\},
\end{eqnarray}
where ``$\cdot$'' refers to element wise multiplications.
%the elements of  $\mathsf{E_3}$, $\mathsf{E_4}$, $\mathsf{E_5}$, $\mathsf{E_6}$, $\mathsf{E_7}$, and $\mathsf{E_8}$
%\begin{equation}
%(E_3)_{ij}=e^{-j(K^z_i-(K^z_j)^*)z};\;\;\;\;\;\; (E_4)_{i}=e^{-jK^z_i(L-z)};  \;\;\;\;\;\; (E_7)_{j}=e^{j(K^z_j)^*z}; 
%\end{equation}
%\begin{equation}
 %(E_6)_{i}=e^{-jK^z_i(L-z)};  \;\;\;\;\;\; (E_7)_{j}=e^{j(K^z_j)^*z};\;\;\;\;\;\; (E_8)_{ij}=e^{-j(K^z_i-(K^z_j)^*)(L-z)}
%\end{equation}

The elements of matrices $\mathsf{K_p}$ and $\mathsf{S}$ are
\begin{equation}\label{K_p_S}
(K_p)_{ij}=\frac{\int(\tilde{\phi}_i^l)^*\tilde{\phi}_i^lds\int(\tilde{\phi}_j^r)^*\tilde{\phi}_j^rds}{\int(\tilde{\phi}_i^l)^*\tilde{\phi}_i^rds(\int(\tilde{\phi}_j^l)^*\tilde{\phi}_j^rds)^*},\;\;\;\;\;\; S_{ij}=\frac{\int\tilde{\phi}_i^r(\tilde{\phi}_j^r)^*ds}{\sqrt{\int(\tilde{\phi}_i^r)^*\tilde{\phi}_i^rds\int(\tilde{\phi}_j^r)^*\tilde{\phi}_j^rds}},
\end{equation}
$(K_p)_{ii}$ describe the self-overlap of the left eigenfunction and right eigenfunction of one mode. $S_{ij}$ ($i\ne j$, $S_{ii}=1$) describe the normalized overlap between modes. The two new parameters $\mathsf{K_p}$ and $\mathsf{S}$, which are firstly proposed in acoustics, are the key result of this paper. Two extreme cases are: When the boundary is acoustic rigid, left eigenfunction is equal to the right eigenfunction, $(K_p)_{ii}=1$; Modes are mutual orthogonal in the standard definition (\ref{standard}), $S_{ij}=0$ ($i\ne j$). At optimum impedance, $(K_p)_{ii}=\infty$, left eigenfunction and right eigenfunction are self-orthogonal; $S_{ij}=1$ ($i\ne j$), two modes coalescence. In all other lining impedance cases, $1\le (K_p)_{ii}<\infty$ and $0\le S_{ij}\le1$ ($i\ne j$). From eq. (\ref{sp}), it is shown that  $\mathsf{K_p}$ and $\mathsf{S}$ play essential roles in the sound propagation and attenuation in lined parts.

\section{Examples}

In this section, we will show the roles of the factors $\mathsf{K_p}$ and $\mathsf{S}$ by an example. We assume that the lining impedance is one segment, axially and circumferentially uniform. The lining structure is made by a resistive facing-sheet separated by one or more honeycomb layers, with the overall panel being backed by a reflective solid backing-sheet. It can be expressed as 
\begin{equation}\label{o_i}
Z/\rho c=R_l+j[0.01+ 0.0020356896K/R-0.20356896*\cot(0.03*K/R)].
\end{equation}
\begin{figure}[htbp]
\begin{minipage}{0.48\textwidth}
\includegraphics[width=\textwidth]{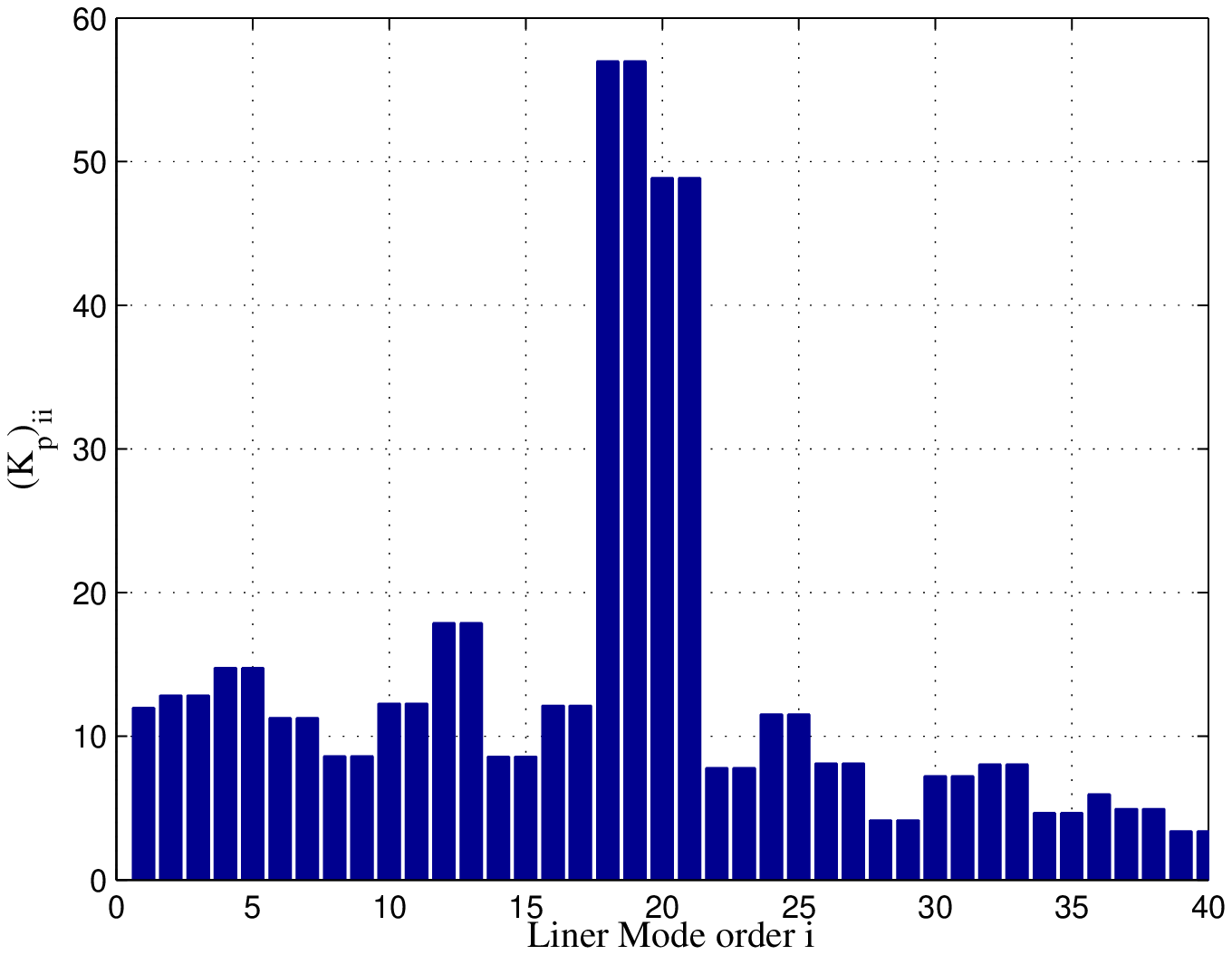}
\caption{\label{K1} $K_{ii}$ for $R_l=1.26$. $i=18-21$ corresponds to the liner modes (9, 1) cosine component, (9, 1) sine component, (9, 0) cosine component, (9, 0) sine component.}
\end{minipage}
\hfill
\begin{minipage}{0.48\textwidth}
\includegraphics[width=\textwidth]{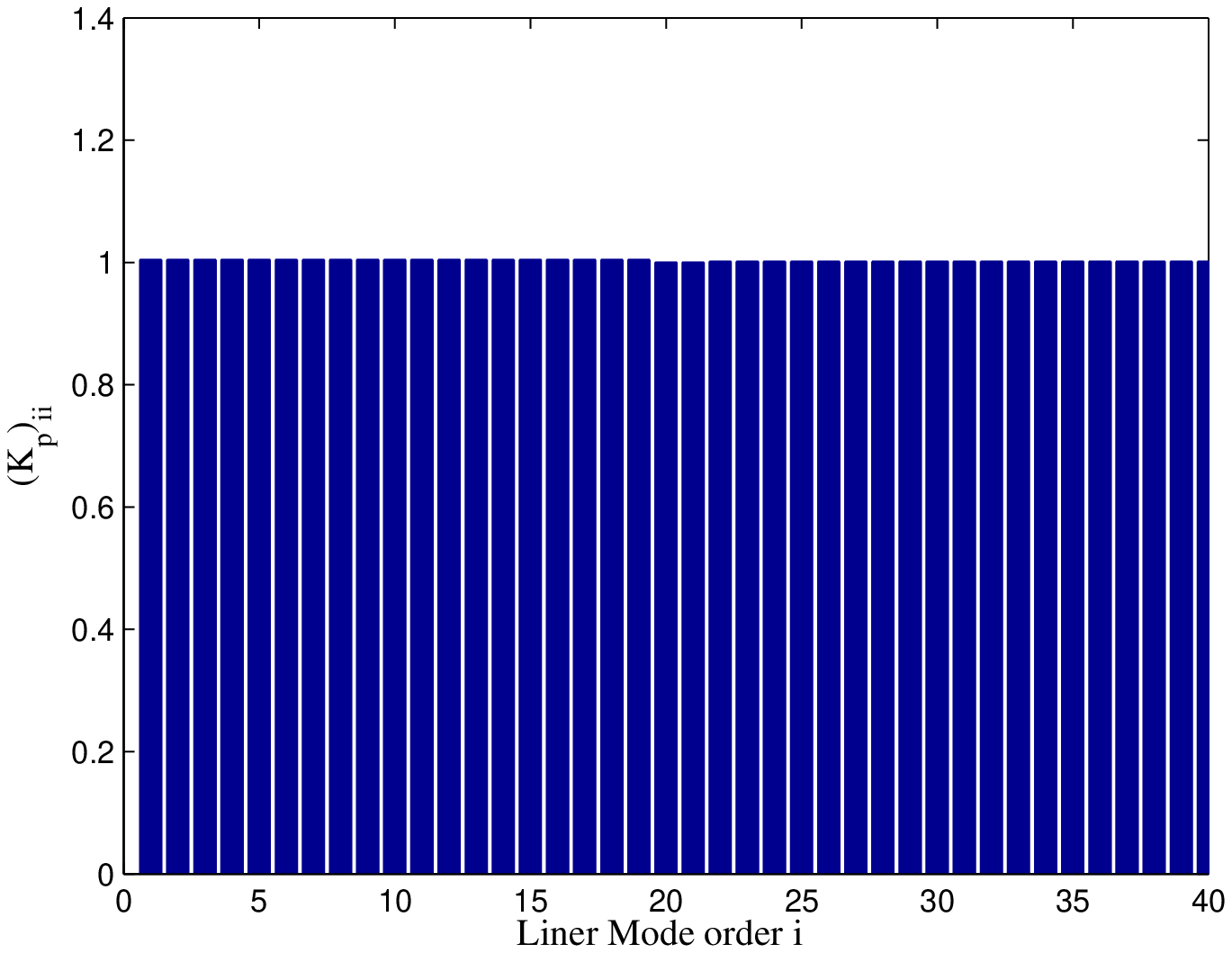}
\caption{\label{K2} $K_{ii}$ for $R_l=0.02$. $i=18-21$ corresponds to the liner modes (9, 0) cosine component, (9, 0) sine component, (9, 1) cosine component, (9, 1) sine component.}
\end{minipage}
\end{figure}

\begin{figure}[htbp]
\begin{minipage}{0.48\textwidth}
\includegraphics[width=\textwidth]{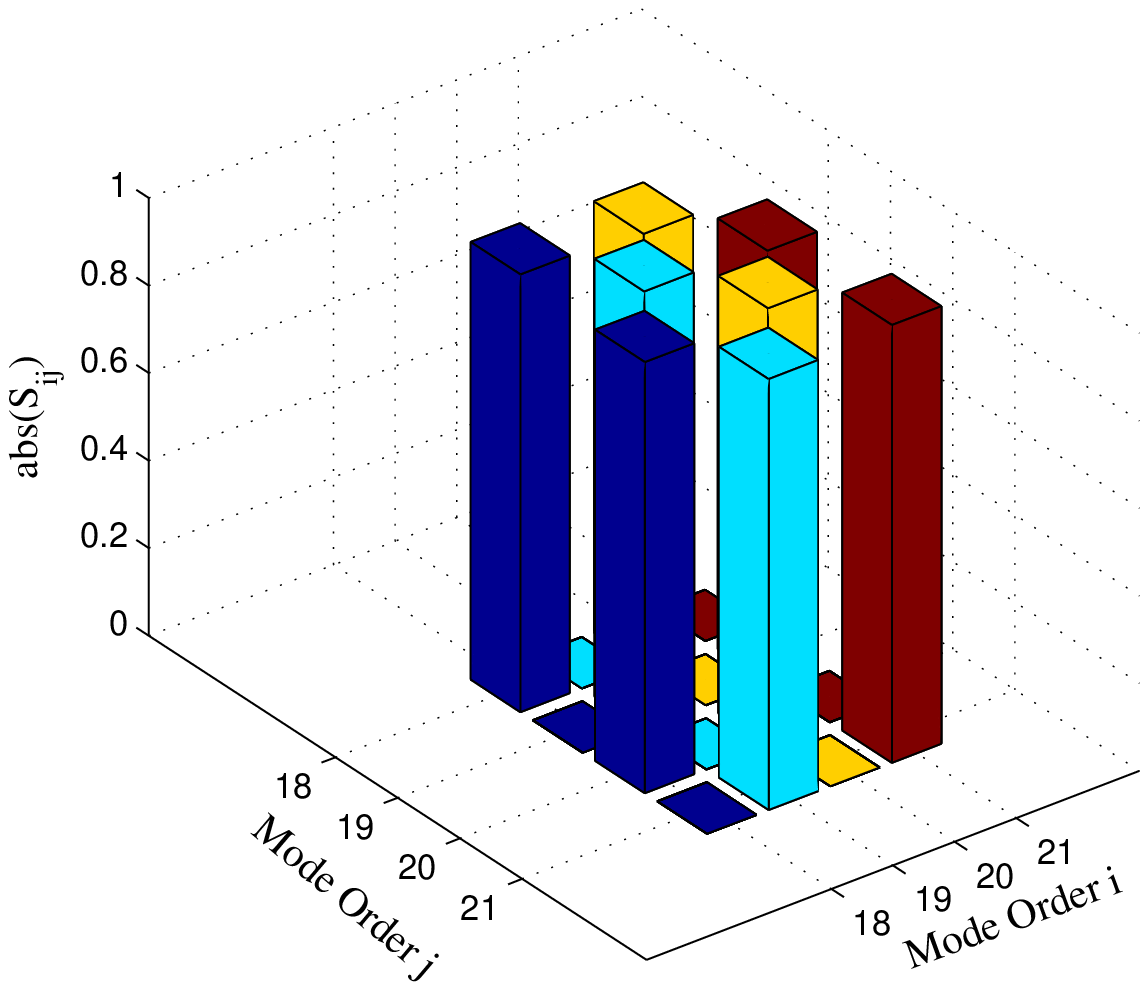}
\caption{\label{S1} $S_{ij}$ for $R_l=1.26$. Mode order $i, j$ is same as in Figs. (\ref{K1}), (\ref{K2}).}
\end{minipage}
\hfill
\begin{minipage}{0.48\textwidth}
\includegraphics[width=\textwidth]{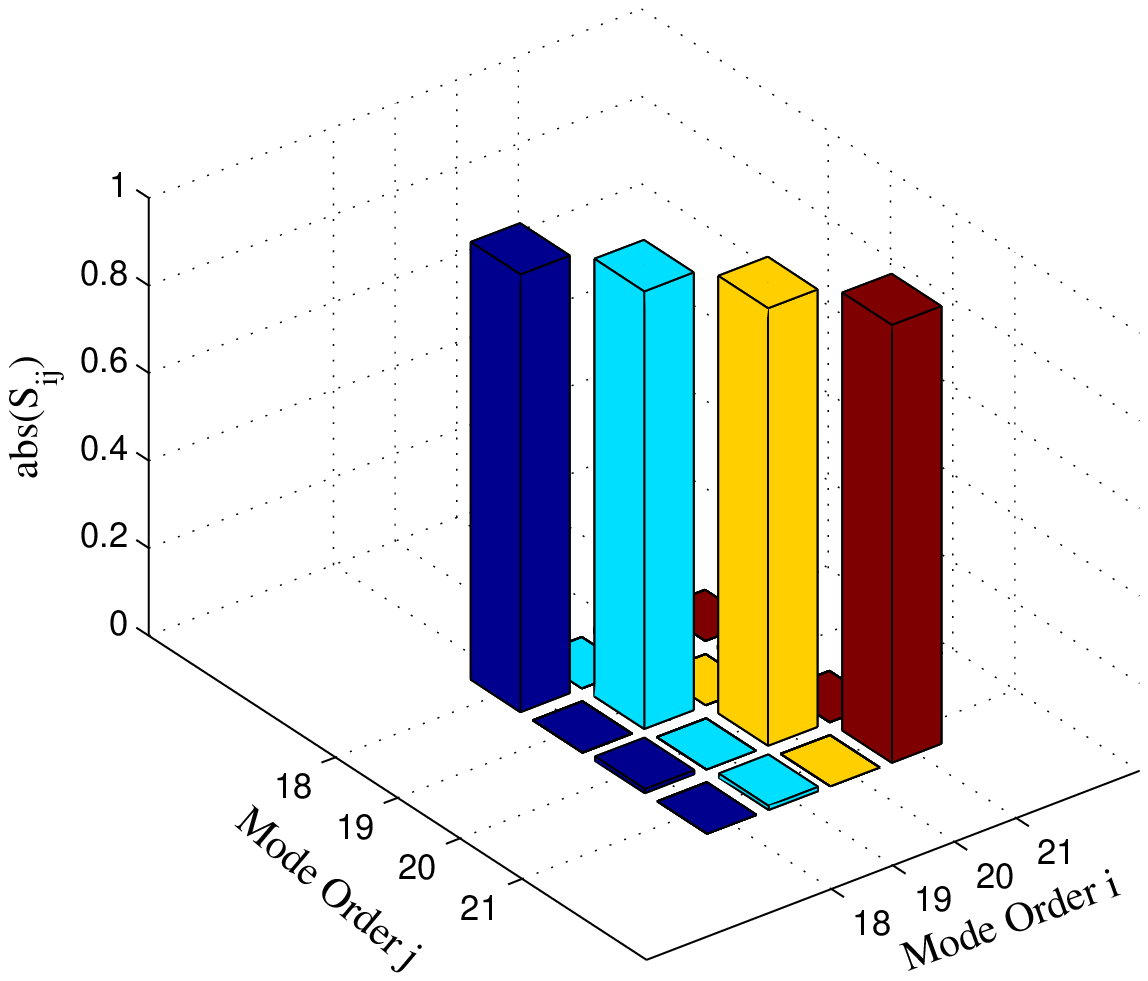}
\caption{\label{S2}  $S_{ii}$ for $R_l=0.02$. Mode order $i, j$ is same as in Figs. (\ref{K1}), (\ref{K2}).}
\end{minipage}
\end{figure}
\begin{figure}[htbp]
\begin{minipage}{0.48\textwidth}
\includegraphics[width=\textwidth]{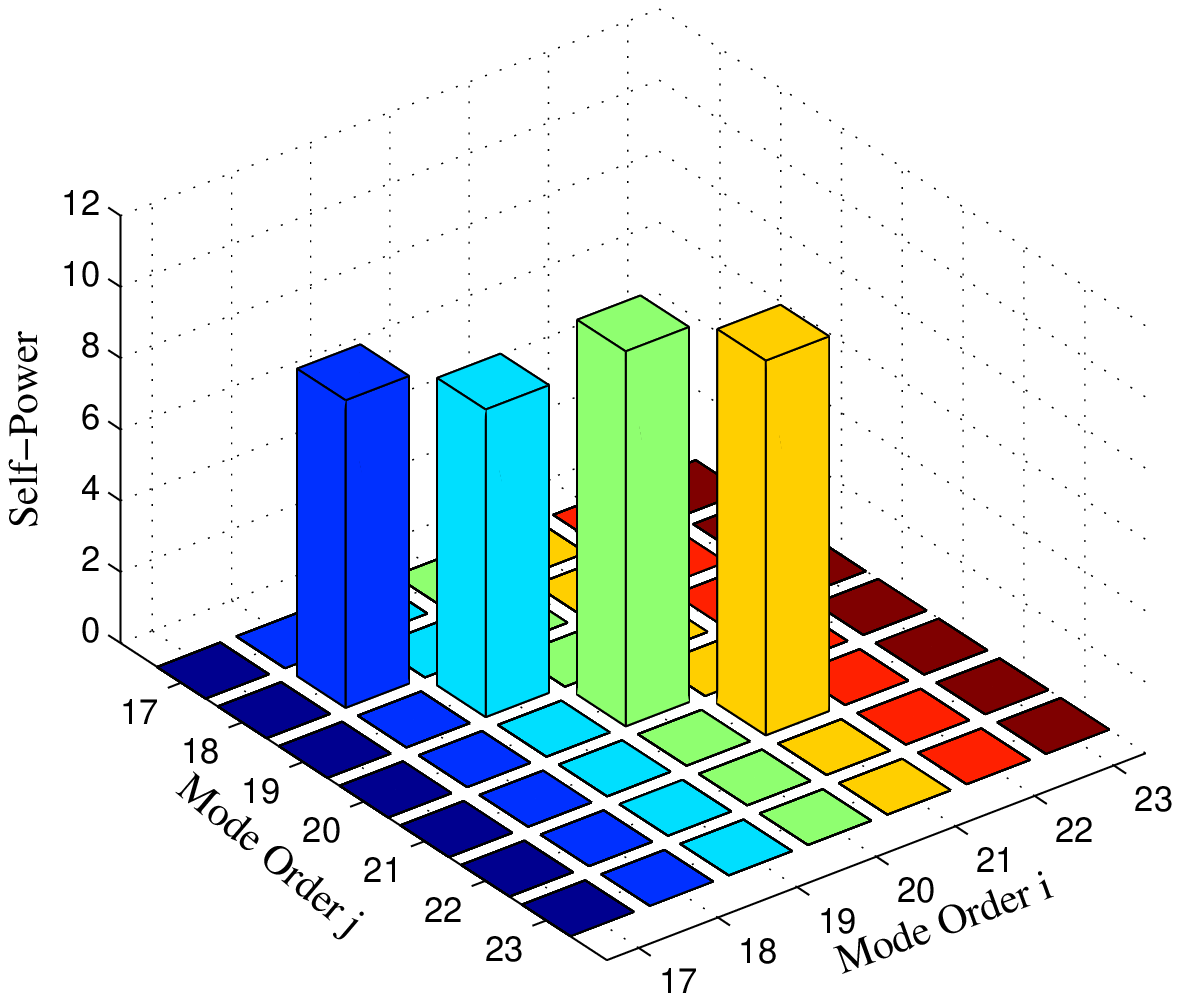}
\caption{\label{sound power} Self-power for $R_l=1.26$. Mode order $i, j$ is same as in Figs. (\ref{K1}), (\ref{K2}).}
\end{minipage}
\hfill
\begin{minipage}{0.48\textwidth}
\includegraphics[width=\textwidth]{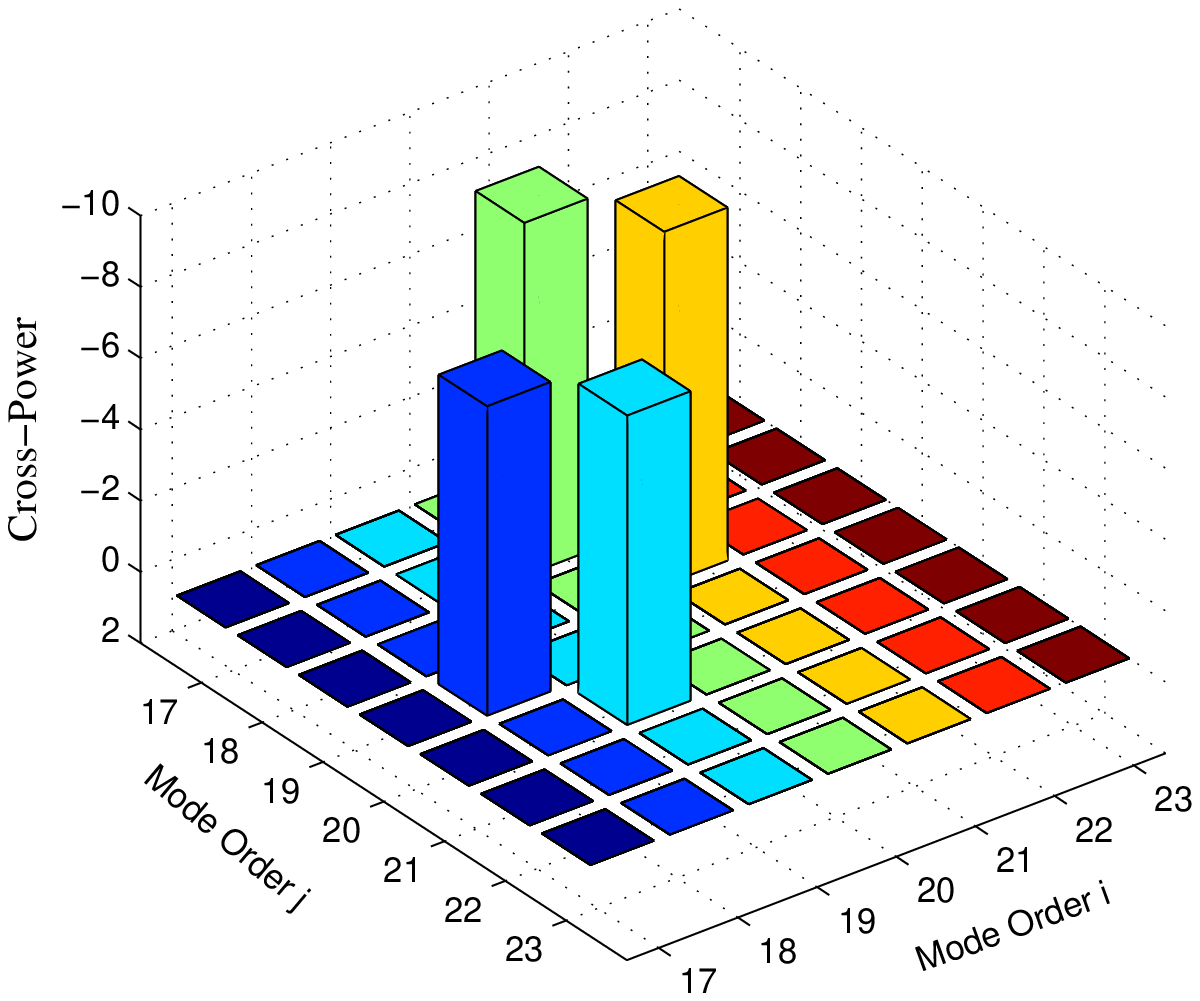}
\caption{\label{sound power1} Cross-power for $R_l=1.26$. Mode order $i, j$ is same as in Figs. (\ref{K1}), (\ref{K2}).}
\end{minipage}
\end{figure}

\begin{figure}[htbp]
\begin{minipage}{0.48\textwidth}
\includegraphics[width=\textwidth]{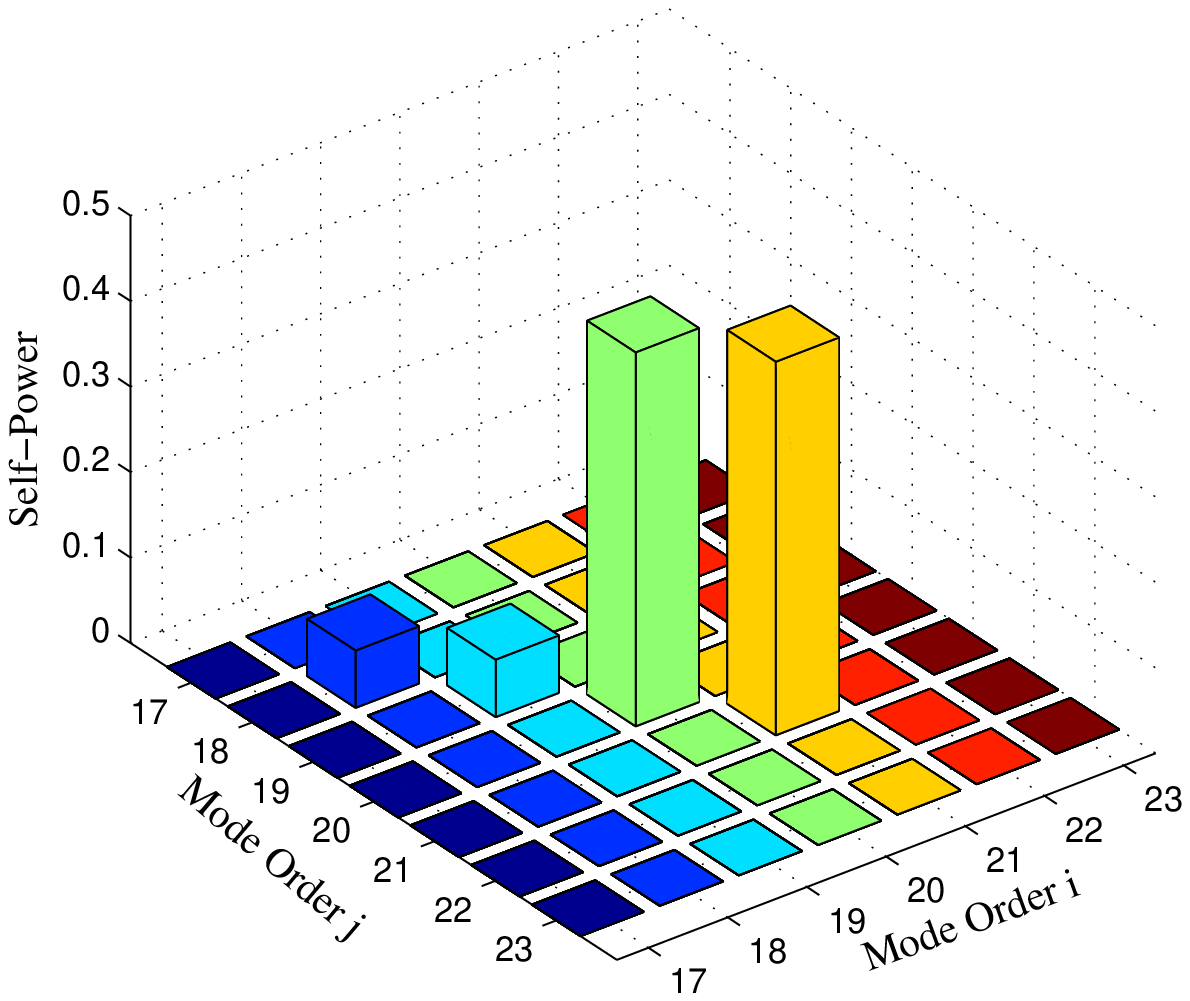}
\caption{\label{sound power2} Self-power for $R_l=0.02$. Mode order $i, j$ is same as in Figs. (\ref{K1}), (\ref{K2}).}
\end{minipage}
\hfill
\begin{minipage}{0.48\textwidth}
\includegraphics[width=\textwidth]{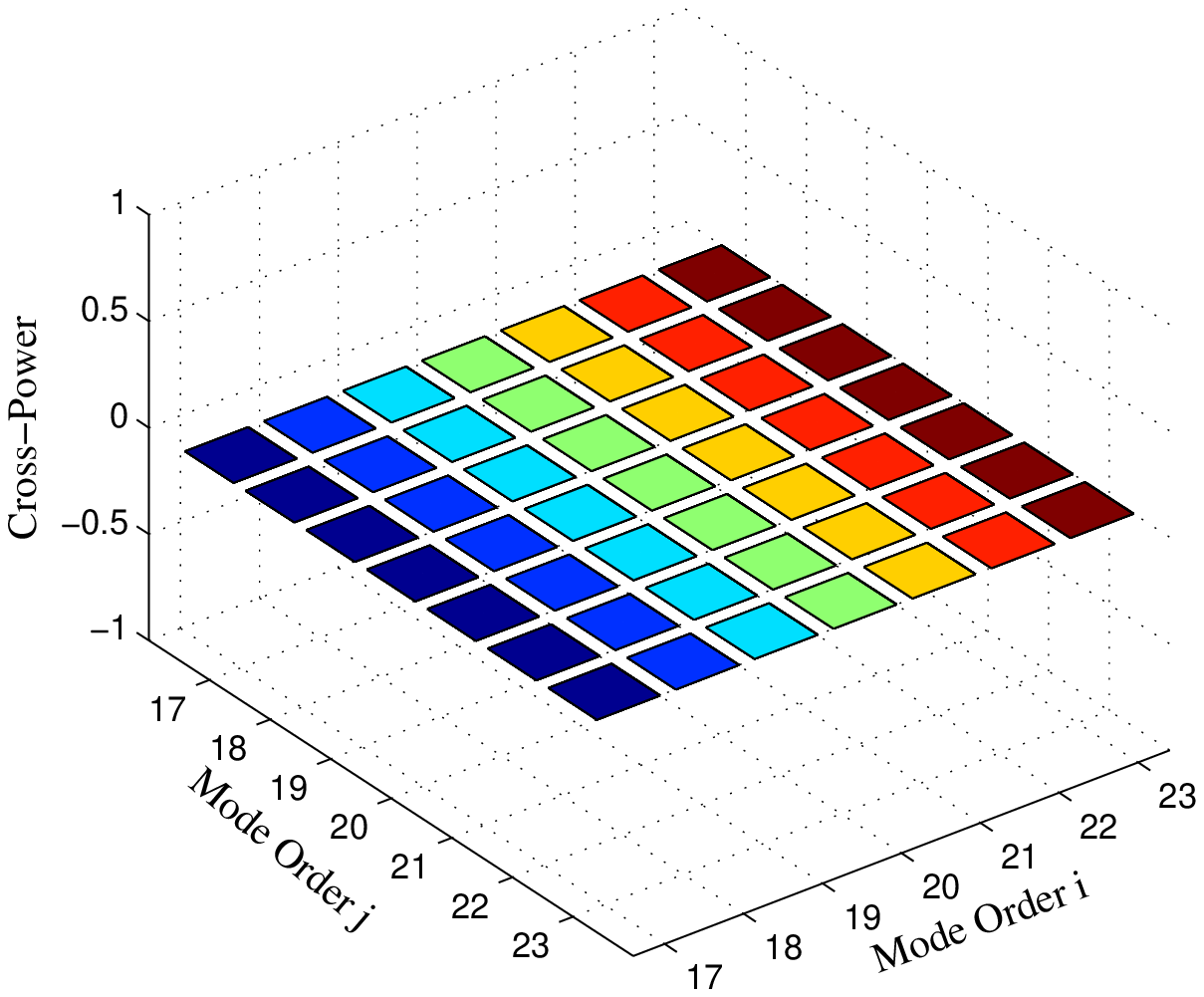}
\caption{\label{sound power3} Cross-power for $R_l=0.02$. Mode order $i, j$ is same as in Figs. (\ref{K1}), (\ref{K2}).}
\end{minipage}
\end{figure}

We will compare two cases $R_l=1.26$ and $R_l=0.02$ for $K=15.05$. In Figs. (\ref{K1}-\ref{sound power3}), we plot $K_{ii}$, $S_{ij}$, self-power and cross-power for liner mode (9, 0) and (9, 1). Mode indices $i=18-21$ refers to the liner modes (9, 1) cosine component, (9, 1) sine component, (9, 0) cosine component, (9, 0) sine component for $R_l=1.26$. Mode indices $i=18-21$ corresponds to the liner modes (9, 0) cosine component, (9, 0) sine component, (9, 1) cosine component, (9, 1) sine component for $R_l=0.02$. 

When $R_l=0.02$, lining impedance is nearly purely reactive including a very small absorption. The left eigenfunctions of modes (9, 0) and (9, 1) are approximately equal to their right eigenfunctions. All $K_{ii}$ are nearly equal to 1 as shown by eq. (\ref{K_p_S}) and Fig. (\ref{K2}). Modes (9, 0) and (9, 1) are approximately orthogonal under the standard definition (\ref{standard}), $S_{ii}=1$ and $S_{ij}\approx 0$ when $i\ne j$. The total sound power is approximately equal to the sum of the each self-power, i. e., the sum of the sound power of modes (9, 0) and (9, 1). The cross-power between modes (9, 0) and (9, 1) are approximately equal to zero as shown in Fig. (\ref{sound power3}). 

On the other hand, When $R_l=1.26$, lining impedance is in the vicinity of optimum impedance of modes (9, 0) and (9, 1). The left eigenfunctions of modes (9, 0) and (9, 1) are equal to the complex conjugate of their right eigenfunctions. $K_{(9,0), (9,0)}=49$ and $K_{(9,1), (9,1)}=57$ as shown Fig. (\ref{K1}). Modes (9, 0) and (9, 1) are bi-orthogonal under the modified definition (\ref{modify}). Modes (9, 0) and (9, 1) are highly similar in the sense of eq. (\ref{K_p_S}). $S_{ij}\approx 1$, where $i, j$ refer to (9, 0) and (9, 1) respectively. The total sound power is not equal to the sum of the self-power, i. e., the sound power of modes (9, 0) and (9, 1). The cross-power between modes (9, 0) and (9, 1) are very important as shown in Fig. (\ref{sound power1}) (attention: in Fig. (\ref{sound power1}), the cross-power is negative). The negative cross-power highly destroy the effects of optimum attenuation of each mode.

\section{Conclusions}

In this paper, we propose two new  physical quantities $\mathsf{K_p}$ and $\mathsf{S}$. $(K_p)_{ii}$ describe the self-overlap of the left eigenfunction and right eigenfunction of one mode. $S_{ij}$ ($i\ne j$, $S_{ii}=1$) describe the normalized overlap between modes. The two new physical quantities describe totally the mode behavior in lined ducts.

In the two extreme cases are: When the boundary is acoustic rigid, left eigenfunction is equal to the right eigenfunction, $(K_p)_{ii}=1$; Modes are mutual orthogonal in the standard definition (\ref{standard}), $S_{ij}=0$ ($i\ne j$). At optimum impedance, $(K_p)_{ii}=\infty$, left eigenfunction and right eigenfunction are self-orthogonal; $S_{ij}=1$ ($i\ne j$), two modes coalescence. In all other lining impedance cases, $1\le (K_p)_{ii}<\infty$ and $0\le S_{ij}\le1$ ($i\ne j$). 

We have shown that $\mathsf{K_p}$ and $\mathsf{S}$ play essential roles in the sound propagation and attenuation in lined parts. We believe that our results are important in understanding the optimum design of liners.


\end{document}